**ANÁLISE DA ATIVIDADE NUCLEAR DE DEZ GALÁXIAS COM ANEL POLAR**

**ANALYSIS OF NUCLEAR ACTIVITY OF TEN POLAR RING GALAXIES**

**Priscila Freitas-Lemes**[1]
**Irapuan Rodrigues**[2]
**Oli Dors**[2]
**Max Faúndez-Abans**[3]

**RESUMO:** *O acúmulo de massa, proveniente do processo de interação que forma as galáxias com anel polar, é um fator que favorece as condições necessárias para disparar atividades nucleares não térmicas. Esse fato fomentou a análise química de dez galáxias com anel polar. Com o objetivo de verificar a presença de núcleo ativo nessas galáxias, construímos diagramas de diagnósticos usando as linhas de Hβ, [OIII], [OI], Hα, [NII] e [SII] e as classificamos quanto ao tipo de atividade nuclear. Para as galáxias que não apresentam choque, determinamos, também, o parâmetro N2 e O3N2. Dessa amostra, identificamos sete galáxias com núcleo ativo e três que se comportam como regiões HII. Uma das galáxias com núcleo ativo foi classificada como Seyfert. Apesar dos nossos dados não fornecerem uma estatística significativa, podemos especular que as galáxias com anel polar são um cenário propício para disparar atividades nucleares não térmicas.*
**PALAVRAS-CHAVE:** galáxias em interação; abundância química de galáxias.

**ABSTRACT:** *The accumulation of mass from the interaction process that forms the polar ring galaxies is a factor that favors the conditions necessary to trigger nonthermal nuclear activities.. This fact encouraged the chemical analysis of ten polar ring galaxies. In order to verify the presence of an active nucleus in these galaxias, we built diagnostic diagrams using lines Hβ, [OIII], [HI], Hα, [NII], and [SII] and classified the type of nuclear activity. For galaxies that do not show shock, the parameters N2 and O3N2 were also determined. From this sample, we identified seven galaxies with an active nucleus and three that behave as HII regions. One galaxy with an active nucleus was classified as Seyfert. Although our data do not provide a statistically significant sample, we can speculate that polar ring galaxies are a setting conducive to trigger non-thermal nuclear activities.*

[1] Doutoranda em Física e Astronomia - Universidade do Vale do Paraíba - UNIVAP / Instituto de Pesquisa e Desenvolvimento - IP&D. E-mail: priscila@univap.br.
[2] Docente da Univap / IP&D. E-mails: irapuan@uivap.br; olidors@univap.br.
[3] Ministério da Ciência, Tecnologia e Inovação - MCTI / Laboratório Nacional de Astrofísica - LNA. E-mail: mfaundez@lna.br.

## 1. INTRODUÇÃO

Ao se estudar o processo de formação e evolução de galáxias ou sistemas estelares, a análise espectral é uma ferramenta. Este estudo é baseado em espectros observados, de onde podemos extrair informações sobre a idade, a metalicidade e pistas de como decorreu o seu processo de evolução. A história química da evolução de uma galáxia é regida por diversas populações estelares que se formaram e evoluíram no decorrer da vida da galáxia, assim, a compreensão da evolução química de uma galáxia pode ser entendida como o desmembramento dessas histórias.

Dentro de uma galáxia, encontramos estrelas de diferentes massas com uma ampla diferença de idade. Isso ocorre pelo fato da população estelar evoluir naturalmente, mesmo quando estudamos galáxias isoladas. Contudo, a interação entre galáxias, que é um evento comum, influencia a evolução da população da galáxia através do acréscimo e compressão de gás, da fusão com outros corpos celestes etc. Em contrapartida, a população estelar pode ser importante para o enriquecimento da população vizinha, que pode ocorrer através de jatos provenientes da região nuclear ou explosões de supernovas.

Neste trabalho, analisaremos a presença de núcleos ativos em dez galáxias com anel polar (veja Tabela 1). As galáxias com anel polar compõem um grupo especial das galáxias aneladas, onde um anel de gás, poeira e estrelas orbita num plano aproximadamente perpendicular ao plano principal da galáxia hospedeira. Por ser um sistema formado através do processo de interação, as PRGs têm uma forte tendência a possuir um núcleo ativo, impulsionado pela transferência de gás do processo de interação.

## 2. CLASSIFICAÇÃO ESPECTRAL DE GALÁXIAS

Analisando as linhas de emissão presentes nos espectros, Baldwin, Philips e Terlevich (1981) - BPT demonstraram que é possível separar as regiões fotoionizadas por estrelas dos núcleos ativos, por meio de *diagramas de diagnósticos*. O conceito introduzido nos diagramas de diagnósticos é usufruir das diferenças entre as linhas de emissão geradas em diferentes tipos de galáxias.

**Tabela 1 - Informações gerais das galáxias estudadas neste trabalho**

| Objeto | PRC[1] | Velocidade Radial | Magnitude | C. Núcleo |
|---|---|---|---|---|
| AM2020-504 | B-19 | 5006±25km/s | 14.5 | LINER |
| NGC 660 | C-13 | 850±1 km/s | 12,00 | LINER |
| UGC 05101 | C-30 | 11802 km/s | 15,10 | Seyfert |
| NGC 0520 | C-44 | 2281 km/s | 12,24 | T.O – LINER |
| NGC 6286 | C-51 | 5501±16 km/s | 14,06 | LINER |
| AM2040-620 | C-56 | 3335±19 km/s | 15.75 | HII galaxy |
| NGC 3310 | D-15 | 993 km/s | 11,15 | HII galaxy |
| UGC 08387 | D-25 | 6985 km/s | 14,70 | T.O – LINER |
| NGC 6240 | D-28 | 7339 km/s | 13,80 | LINER |
| HRG54103 |  | 6444±23km/s | 14.38 | Seyfert |

[1] **Classificação proposta por Whitmore *et al*. (1990).**

## 3. ATIVIDADE NUCLEAR EM GALÁXIAS COM ANEL POLAR

No mecanismo de formação de galáxias com anel polar, assim como em todo processo de interação de galáxias, a transferência de gás (massa) de uma componente a outra varia de sistema para sistema, desde acréscimo de gás e estrelas à fusão total. Sendo assim, com o acúmulo de massa, é razoável supor que as PRGs apresentam as condições adequadas para disparar atividades nucleares não térmicas. Reshetnikov, Faúndez-Abans e de Oliveira-Abans (2001) determinaram que a fração de galáxias ativas é alta nas PRGs e, também, em possíveis candidatas a PRGs: sendo de 13-25% para Seyfert e 33-41% LINERs. Esses resultados, porém, podem ter deixado de fora uma grande quantidade de objetos, já que foram derivados de observações no visível, o que, possivelmente, mascarou um grande número PRGs que contenham um AGN, já que a região central de uma grande quantidade de galáxias aneladas está envolvida em um casulo de poeira.

Analisando os espectros dessas PRGs, foi possível mensurar a contribuição de Hβ, [OIII], [OI], Hα, [NII] e [SII] dos espectros, veja as razões de linhas consideradas, neste trabalho, na Tabela 2. Com esses valores, classificamos as regiões nucleares de cada uma das galáxias em AGNs ou como regiões HII. Para tal classificação, utilizamos o diagrama proposto por BPT (1981), Coziol *et al.* (1999) e Kewley (2006). Contamos com espectros do telescópio Palomar 200in *Double Spectrograph,* cuja cobertura espectral é de 3800 – 8000 Å.

## 4. RESULTADOS

Através da análise dos espectros das PRGs, analisamos a contribuição das linhas Hβ, [OIII], [OI], Hα, [NII] e [SII]. Com esses valores, classificamos o núcleo dessas galáxias em AGNs ou galáxias normais. Para tal classificação, utilizamos o diagrama de diagnósticos, proposto por Coziol *et al.* (1999) e Kewley *et al.* (2001). Esses diagramas relacionam as razões de linhas [OIII]/Hβ [NII]/Hα e [SII]/Hα.

Analisando o diagrama de diagnósticos log([NII]/Hα) x log([SII]/Hα) (Figura 2), verificamos que apenas duas galáxias do nosso grupo não possuem núcleo ativo.

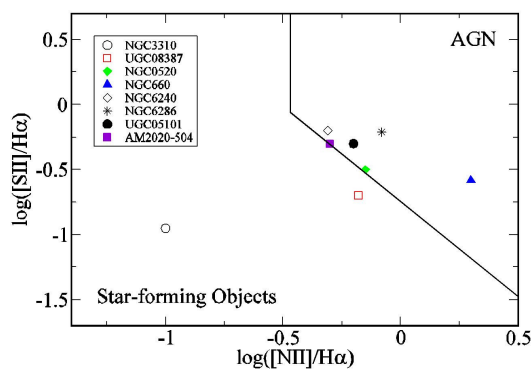

**Figura 2 - Diagrama de diagnóstico log[NII]/Hα *versus* log[SII]/Hα. A linha proposta por Coziol *et al.* (1999) separa os AGNs das regiões HII.**

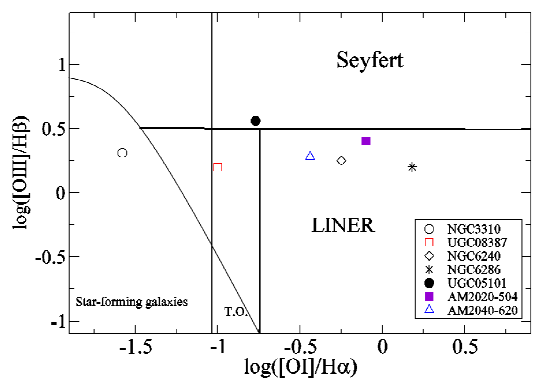

**Figura 3 - Diagrama de diagnóstico log[OI]/Hα *versus* log [OIII]/Hβ. A curva corresponde à separação empírica entre núcleo ativos e galáxias HII (KEWLEY *et al.*, 2001a). A separação entre núcleos ativos LINERs, e Seyferts é dado por Filippenko e Terlevich (1992). Entre as galáxias HII e as LINERs há a separação destinada aos 'objetos de transição' (indicado por T.O.), descritos por Ho, Shields e Filippenko (1993).**

Usando as razões log([OI]/Hα) x log ([OIII]/Hβ) (Figura 3) e a curva proposta por Kewley *et al.* (2001a), confirmamos a classificação de núcleos ativos e regiões fotoionizadas por estrelas, feito no diagrama anterior. Agora, usando as linhas propostas por Filippenko e Terlevich (1992), separamos as galáxias Seyfert dos núcleos ativos de baixa ionização, LINERs. Novamente só a PRG NGC3310 se comporta como uma região HII, já a galáxia UGC08387, nesse diagrama, está na chamada Zona de Transição. Somente a PRG UGC 05101 se comporta como uma Seyfert. Os "objetos transitórios" (do inglês *transition objects* - TO) são apresentados nesse diagrama. Ho, Shields e Filippenko (1993) propuseram que alguns objetos podem estar passando por uma *zona transitória* e podem, futuramente, apresentar um núcleo ativo.

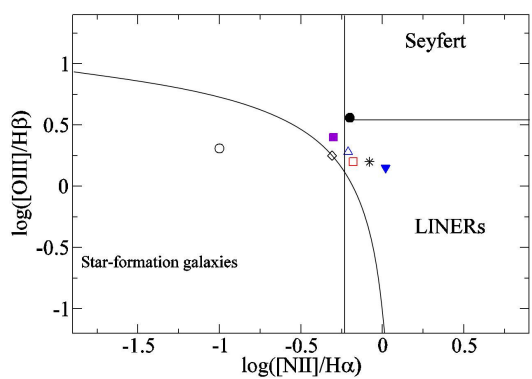

**Figura 4 - Diagrama de diagnóstico log [NII]/Hα *versus* log[OIII]/Hβ. A curva separa as galáxias com núcleos ativos das galáxias HII, como proposto em Kewley *et al.* (2001a). Kauffmann *et al.* (2003) propuseram a linha que separa as galáxias Seyfert das LINERs, usadas neste diagrama.**

Na Figura 4, mostramos o diagrama de diagnóstico que relaciona as razões log ([NII]/Hα) *x* log([OIII]/Hβ). Nesse diagrama, também fica evidente que a galáxia PRG NGC 3310 se comporta como uma região HII, assim como foi verificado nos dois diagramas anteriores. Kauffmann *et al.* (2003) propuseram a linha que separa as galáxias com núcleo ativo, das galáxias Seyfert e das LINERs. Nessa classificação, a galáxia PRG UGC05101 é a única a ser classificada como Seyfert, assim como na Figura 3. A galáxia UGC08387, nesse diagrama, é classificada como um núcleo ativo de baixa ionização, LINER.

## 5. CONCLUSÕES

Analisando os espectros das PRGs selecionadas, identificamos as linhas de emissão de Hβ, [OIII], [OI], Hα, [NII] e [SII]. Usando razões de linhas já consagradas e separações empíricas, classificamos o núcleo dessas galáxias em AGNs ou regiões HII.

Analisamos os diagramas de diagnóstico usando as linhas propostas por Baldwin, Phillips e Tervelick (1981), Ho, Shields e Filippenko (1993), Coziol *et al.* (1999), Kewley *et al.* (2001) e Kauffmann *et al.* (2003). Apenas a galáxia PRG NGC3310 não tem núcleo ativo.

A galáxia PRG UGC05101 apresentou atividade nuclear, é uma galáxia Seyfert.

Os nossos dados não fornecem uma estatística significativa dentre as galáxias com anel polar, entretanto, quando observamos os resultados encontrados em Reshetnikov e Combes (1994), Willner *et al.* (1985), van Driel *et al.* (1995) e Reshetnikov, Faúndez-Abans e de Oliveira-Abans (2001), podemos especular que as galáxias com anel polar apresentam condições favoráveis à formação de atividades nucleares não térmicas.

## REFERÊNCIAS